\documentclass[runningheads]{llncs}

\usepackage{graphicx}
\usepackage{url}
\usepackage{color}
\usepackage{hyperref}

\newcommand{\q}[1]{\lq\lq{}{}#1\rq\rq{}{}}
\newcommand{\sq}[1]{\lq{}#1\rq{}}
\newcommand{\system}[0]{{\em Synthesis}}

\begin{document}

\title{
    Towards Semantic Interoperability in Historical Research:  Documenting Research Data and Knowledge with \system
}
\titlerunning{Documenting Research Data and Knowledge with \system}

\author{
	Pavlos Fafalios\inst{1}
	\and 
	Konstantina Konsolaki\inst{1}
	\and  
    Lida Charami\inst{1}
	\and  
	Kostas Petrakis\inst{1}
	\and  
	Manos Paterakis\inst{1}
	\and  
	Dimitris Angelakis\inst{1}
    \and 
    Yannis Tzitzikas\inst{1,2}
    \and 
    Chrysoula Bekiari\inst{1}
    \and 
    Martin Doerr\inst{1}
}

\authorrunning{Pavlos Fafalios et al.}

\institute{
	Centre for Cultural Informatics and Information Systems Laboratory,\\FORTH-ICS, Heraklion, Greece 
	\and 
	Computer Science Department, University of Crete, Heraklion, Greece\\
	\email{\{fafalios, konsolak, lida, cpetrakis, paterakis, agelakis, tzitzik, bekiari, martin\}@ics.forth.gr}
	}

\maketitle

\begin{abstract}
A vast area of research in historical science concerns the documentation and study of artefacts and related evidence. Current practice mostly uses spreadsheets or simple relational databases to organise the information as rows with multiple columns of related attributes.  
This form offers itself for data analysis and scholarly interpretation, however it also poses problems including 
i)~the difficulty for collaborative but controlled documentation by a large number of users, 
ii)~the lack of representation of the details from which the documented relations are inferred,
iii)~the difficulty to extend the underlying data structures as well as to combine and integrate data from multiple and diverse information sources, and 
iv)~the limitation to reuse the data beyond the context of a particular research activity.
To support historians to cope with these problems, in this paper we describe the \system\ documentation system and its use by a large number of historians in the context of an ongoing research project in the field of History of Art. The system is Web-based and collaborative, and makes use of existing standards for information documentation and publication (CIDOC-CRM, RDF), focusing on semantic interoperability and the production of data of high value and long-term validity.

\keywords{Historical Research, Documentation, Digital Humanities, Semantic Interoperability}   
\end{abstract}

\section{Introduction}

Historical science is the field that describes, examines, and questions a sequence of past events, and investigates the patterns of cause and effect that are related to them.
A vast area of research in this field concerns the discovery, collection, organisation, presentation, and interpretation of information about historical events. This includes either the \textit{digitization} (and then \textit{curation}) of archival sources, like in~\cite{petrakis2021,fafalios2021FastCat} for the case of Maritime History, or the detailed \textit{documentation} of cultural artefacts and related evidence~\cite{bekiari2015}, with the latter being the focus of this paper.

Although computing in historical research has developed enormously over the last years, with Semantic Web technologies starting playing a significant and ever increasing role \cite{merono2015semantic}, 
information management problems still exist and are still vast and very varied.
Current practice mostly uses spreadsheets or simple relational databases to organise the information as rows with multiple columns of related attributes.\footnote{We have witnessed this through our participation in a large number of projects in the fields of cultural heritage and digital humanities, and our collaboration with researchers in these fields.} This form offers itself for data analysis and scholarly interpretation, however it also poses problems including 
i)~the difficulty for collaborative but controlled documentation by a large number of historians of different research groups, 
ii)~the lack of representation of the details from which the documented relations are inferred, important for the long-term validity of the research results,
iii)~the difficulty to combine and integrate information extracted from multiple and diverse information sources documented by more than one researcher, 
iv)~the difficulty to easily extend the existing data structures on demand for enabling the incorporation of additional information of historical interest (not originally thought),
v)~the difficulty of third parties to understand and re-use the documented data, resulting in the production of data with limited longevity that lacks semantic interoperability.

To try coping with these problems, in this paper we present the \system\ documentation system and its use by a large number of historians in the context of a European research project (ERC) of History of Art, called RICONTRANS. 
\system\ utilises XML technology, offering flexibility in terms of versioning, workflow management and data model extension, and focuses on semantic interoperability by making use of existing standards for data modelling and publication, in particular the formal ontology (ISO standard) CIDOC-CRM \cite{Doerr_2003} and RDF. 
The aim is the production of data with high value, longevity, and long-term validity that can be (re)used beyond a particular research activity.

We show how the documentation process is performed by researchers working in the RICONTRANS project, the data model used, the functionality and user interface offered by \system, as well as how the documented data is transformed to a rich semantic network of linked data (an RDF knowledge graph). We also discuss lessons learned from our collaboration with historians of RICONTRANS as well as future work related to data dissemination and exploitation.

The rest of this paper is organised as follows: 
Section~\ref{sec:background} describes the context of this work, the requirements and the corresponding challenges.
Section~\ref{sec:dataMngment} details how the \system\ system is used for data documentation in historical research. 
Section~\ref{sec:UI} presents the user interface of \system\ and provides usage statistics.
Section~\ref{sec:lessonsLearnedAndFuture} discusses lessons learned and future data exploitation. Finally, Section~\ref{sec:conclusion} concludes the paper and discusses interesting directions for future work.

\section{Context, Requirements and Challenges}
\label{sec:background}

\subsection{The RICONTRANS Project}

RICONTRANS\footnote{RICONTRANS - \textit{Visual Culture, Piety and Propaganda: Transfer and Reception of Russian Religious Art in the Balkans and the Eastern Mediterranean (16th - early 20th c.)}. ERC Consolidator Grant (ID: 818791). 1 May 2019 - 30 April 2024. \url{https://ricontrans-project.eu/}} 
is an ongoing European research project in the field of History of Art consisting of research groups in Greece, Serbia, Romania, Bulgaria, and Russia \cite{dumitran2019ricontrans}.
The project investigates the transnational phenomenon of artefact transfer and the various aspects of the reception of these objects in the host societies, in different historical periods and circumstances.
The focus is on Russian religious artefacts brought to the Balkans the period 16th-20th centuries, which are now preserved in monasteries, churches or museum collections.

In particular, the project aims to
i) map the phenomenon in its long history by identifying preserved objects in the region;
ii) follow the paths through which these art objects were brought to the Balkans and the Eastern Mediterranean;
iii) identify and classify the mediums of their transfer; 
iv) analyse the dynamics and the various moving factors (religious, political, ideological) of this process; 
v) study, analyse and classify these objects according to their iconographic and artistic particularities; 
vi) inquire into the aesthetic, ideological, political, and social factors which shaped the context of the reception of the transferred objects in the various social, cultural and religious environments; and 
vii) investigate their influence on the visual culture of the host societies.

\subsection{Data Management Requirements and Challenges}
\label{subsec:reqs}
To achieve these objectives, art historians and other researchers of RICONTRANS first need to collect primary and secondary sources, such as archival sources, old books and newspapers, oral history sources, that provide information about Russian artefacts and their transfers to the Balkans. 
The collected information, as well as the knowledge derived by the analysis of the sources, must then be documented in detail and stored in a database in a form that allows its effective exploitation for both current and future research. 

Specifically, the database should contain information about \textit{art objects} (such as icons, triptychs, crosses and censers), \textit{object transfers} (from/to location, purpose of transfer, etc.), \textit{historical figures} (involved in transfers), \textit{locations} (such as cities, villages, monasteries, churches and museums), as well as \textit{related events} (such as the ordination of archbishop, or the erection of a church). 
It must also provide metadata information about the collected \textit{sources}, since this is important for tracking provenance information about the research findings (and thus ensure their long-term validity). Finally, it must allow including (and documenting) \textit{digital files} such as images of art objects, or scans of documents. 

To enable the construction of such a database, we need to cope with the below main data management challenges:
\begin{itemize}
    \item How to support \textit{collaborative} data entry, documentation and curation by a large number of researchers belonging to different research teams that are spread across the world? How to provide to all researchers a common and secure place for storing and accessing their data internally and releasing parts of it to a wider audience when they want to do so?
    \item How to balance between documentation \textit{richness} and database \textit{usability}? How to support researchers in providing detailed information about the documented entities, as well as additional metadata/provenance information, in a structured but straightforward way? 
    \item How to facilitate easy \textit{extension} of the database schema for allowing documenting new type of information about the documented entities? In ongoing research projects where new data sources might become available at any time, frequent updates of the database schema are unavoidable.
    \item How to \textit{control} the data entry process for certain pieces of documented information so that a \textit{common terminology} is used across researchers performing the documentation? This is very important for enabling effective information integration and data exploration. 
    \item How to facilitate \textit{future exploitation/reuse} of the database by others (beyond the particular research project)? How to enable easy integration of the data with other relevant data provided by other researchers? How to ensure the long-term validity and longevity of the data? 
\end{itemize}


\section{Data Documentation with \system}
\label{sec:dataMngment}

We provide an overview of the system (Sect.~\ref{subsec:overview}),  detail its data model and the supported types of documentation fields (Sect.~\ref{subsec:dataModel}), and discuss how the data is transformed to a semantic network, i.e., an RDF knowledge graph (Sect.~\ref{subsec:transformation}).

\subsection{System Overview}
\label{subsec:overview}

\system\ is a Web-based system for the collaborative documentation of information and knowledge in the fields of cultural heritage and digital humanities. 
It utilises XML technology and a multi-layer architecture, offering high \textit{flexibility} and \textit{extensibility} (in terms of data structures and data types), as well as \textit{sustainability} (each documented entity, such as an object or object transfer, is stored as an XML document readable by both humans and machines). Its database server is \textit{eXist-db}\footnote{\url{http://exist-db.org/}}, a native XML database.
Also, \system\ is \textit{multilingual}, supporting the parallel use of multiple languages for documentation, and supports \textit{versioning} of the documented information.

The system supports four roles of users:
i)~\textit{system administrator}, responsible for the whole system, with rights to create new \sq{organisations} (groups); 
ii)~\textit{organisation administrator}, responsible for the documentation process of a particular organisation, with rights to create new editors and guests for this organisation;
iii)~\textit{editor}, belonging to a specific organisation, with rights to create and document entities for this organisation; 
iv)~\textit{guest}, belonging to a specific organisation, with rights to only view the documented entities of a specific organisation.
As we detail below, in \system\ users create and document \textit{entities} belonging to a set of pre-configured \textit{entity types}. Users of role \sq{editor} can only edit entities created by themselves, can provide edit access to other users, and can view only the entities created by editors belonging to the same organisation. However, the management of rights can be easily adjusted for any specific need. For example, one or more editors can be configured to have edit access to all entities because, for instance, they have the responsibility to make corrections. 

\system\ has embedded processes for transforming the data stored in the XML documents to an ontology-based RDF dataset (knowledge graph), thus supporting the creation of a knowledge base (KB) of integrated data. 
Contrary to approaches that support users in creating a KB from the beginning, such as ResearchSpace \cite{oldman2018reshaping} or WissKi \cite{scholz2012wisski}, 
\system\ decouples data entry (made by the research team) from the ontology-based integration and creation of the KB (a process supported by data engineers). 
The main reasons behind this decision are the following (inspired by \cite{doerr2008dream}): 
\begin{itemize}
    \item Versioning in a KB is difficult; individual contributions,  alternatives, corrections, etc., all in the same \textit{pool of valid knowledge} can hardly be regarded as a standard procedure. We consider a KB as an ideal tool for integrating the \textit{latest stage of knowledge} acquired through diverse processes. 
    \item We regard as very different a KB of facts believed together as true, versus managing, coordinating and consolidating the knowledge acquisition process of a large research team. This requires a document structure such as XML, for making local versioning,  workflow management, provenance tracing, and exchanging documents between team members easy.
    \item Decoupling data entry from KB creation allows the straightforward production of different KB versions, considering different ontologies or different versions of the same ontology. This only requires creating and maintaining the schema mappings that transform the documented data to RDF.
\end{itemize}

\subsection{Data Model}
\label{subsec:dataModel}

The data model used by \system\ is carefully designed for a given application domain (History of Art, in our case), with a particular focus on  \textit{semantic interoperability} \cite{ouksel1999semantic,Doerr_2003}. 
This notion is defined as the ability of computer systems to exchange data with unambiguous, shared meaning. 
\system\ achieves this by 
a)~linking each element of its data model to a domain ontology, 
b)~allowing users to add metadata about the data, and 
c)~allowing users to link a term to a controlled (shared) vocabulary or thesaurus of terms (more below).

\begin{figure}[t]
    \vspace{-3mm}
    \centering
    \includegraphics[width=10.7cm]{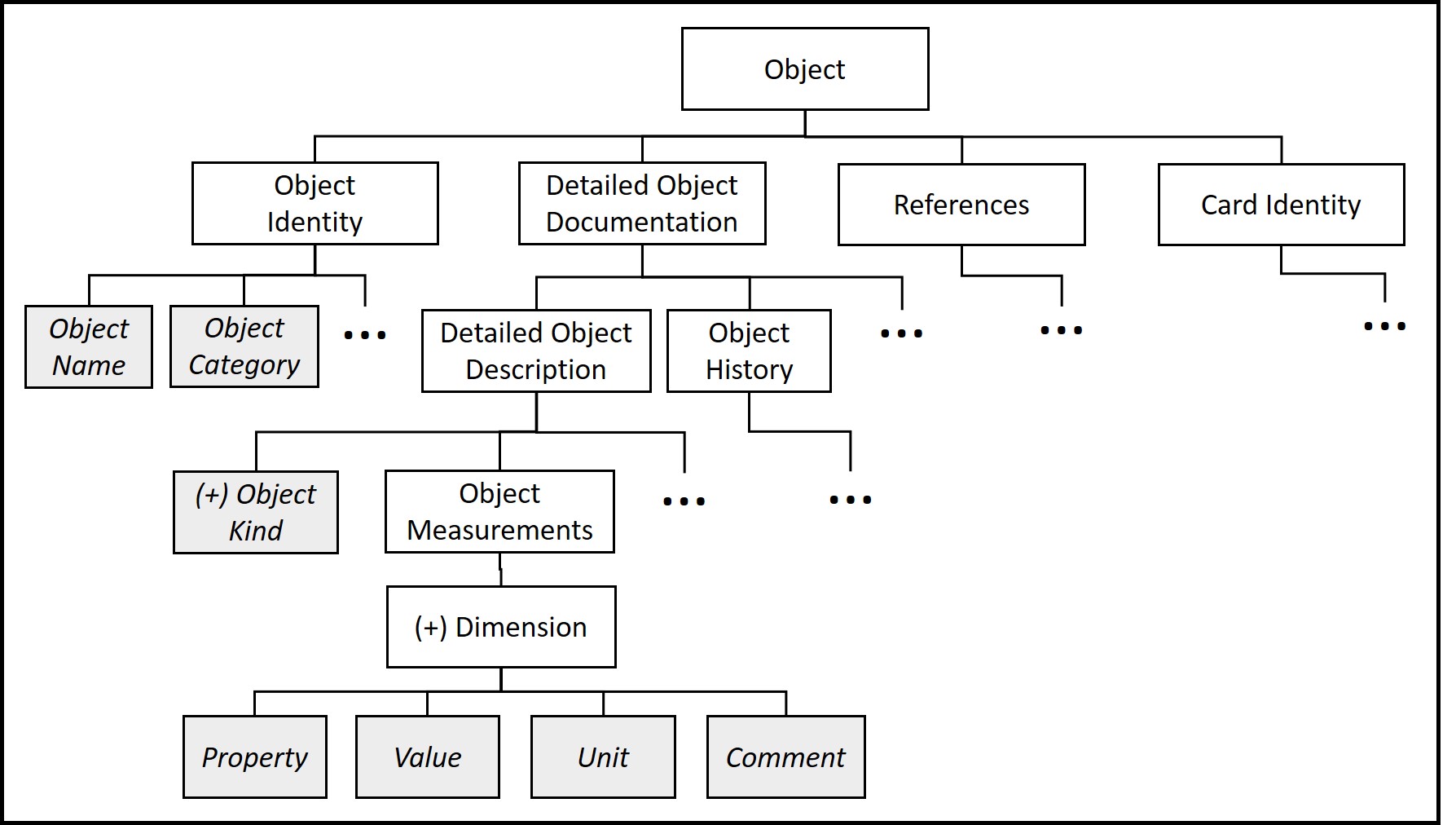} 
    \vspace{-3mm}
    \caption{A part of the schema of the entity type {\tt Object}.}
    \label{fig:schemaExample}
    \vspace{-4mm}
\end{figure}

A user in \system\ can create and start documenting \textit{entities} organised in \textit{entity types}. Each \textit{entity type} has its own data structure (\textit{schema}). A \textit{schema} is XML-based, containing a set of \textit{fields} organised in an hierarchical (tree-like) structure. The leaves in this tree-like structure are the \textit{documentation fields} that are to be filled by the users. Fig.~\ref{fig:schemaExample} shows a small part of the schema of the entity type \sq{Object}, as configured for the RICONTRANS project.

\begin{figure}
    \vspace{-3mm}
    \centering
    \includegraphics[width=10.7cm]{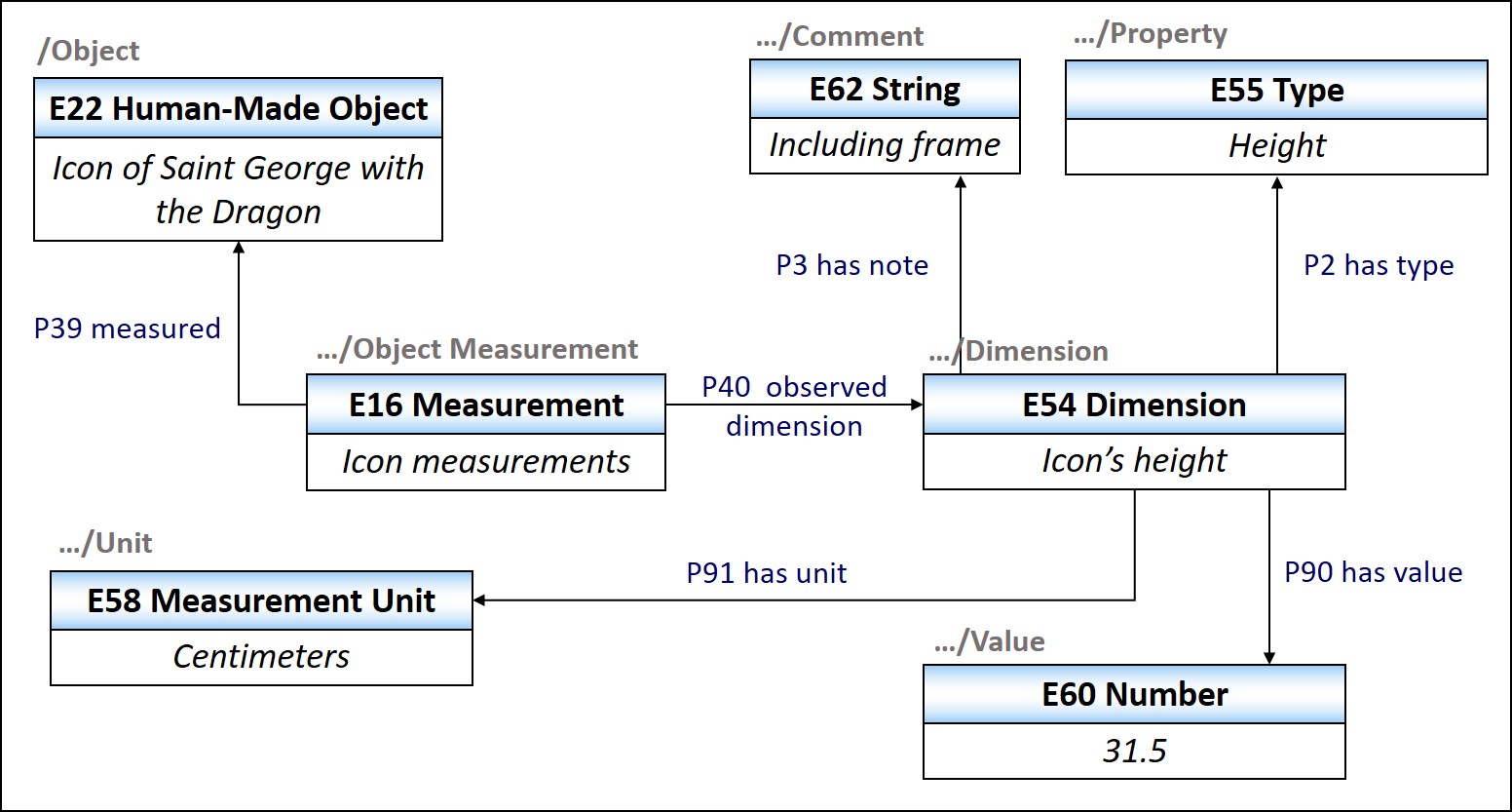} 
    \vspace{-3mm}
    \caption{Mapping of Object Measurement to CIDOC-CRM.}
    \label{fig:crmMappingExample}
\end{figure}

The schema of each entity type is carefully designed to be fully compatible with the CIDOC Conceptual Reference Model (CRM).\footnote{\url{http://www.cidoc-crm.org/}, \url{https://www.iso.org/standard/57832.html}} 
CIDOC-CRM is a high-level, event-centric ontology (ISO standard) of human activity, things and events happening in spacetime, providing definitions and a formal structure for describing the implicit and explicit concepts and relationships used in cultural heritage documentation~\cite{Doerr_2003}. This means that there is a mapping between the schema of an entity type and CIDOC-CRM, allowing the straightforward transformation of the data to a semantic network (RDF graph) that is compatible with \mbox{CIDOC-CRM}. 
For example, Fig. \ref{fig:crmMappingExample} shows how an \textit{Object Measurement} (as documented for an object; cf.~Fig.~\ref{fig:schemaExample}) is mapped to CIDOC-CRM.

In RICONTRANS, the following entity types are available for documentation, each one having its own XML schema. Each schema was designed after extensive discussions with the historians of the RICONTRANS project and has been updated several times in order to allow documenting additional information, not originally thought.  
\vspace{-1mm}
\begin{itemize}
        \item \textit{Objects}: The documentation fields are organised in the following categories: Object Identity (indicative fields: code, name, originator of reference, collection, category, basic material(s), main object image),
        Detailed Object Description (indicative fields: other object names, measurements, object decoration, inscriptions, stamps, locations, photographic documentation),
        Object History (historical events, use, acquisition),
        References (source references, bibliographic references, other related materials),
        Card Identity (scientific supervisor, scientific associates). 
        
        \item \textit{Object Transfers}: Allows documenting information about transfers of objects. Indicative fields:  transfer name/title, transfer date, transferred object, from location, to location, description, transfer purpose, per\-son(s) involved, based on (link to source(s), source passage(s), bibliography).
        
        \item \textit{Routes}: Allows grouping a set of object transfers based on a particular object, type of objects, or other criteria. Indicative fields: route name, object transfers, creation information (author, date). 

        \item \textit{Archival Sources}: Allows documenting information about archival sources used to obtain information about an entity of interest (e.g., about an object, object transfer, historical figure, etc.). Indicative fields: title, subject area, short description, category, type, collection, series, file, language.
        
        \item \textit{Books}: Allows documenting information about (old) books used to obtain information about an entity of interest. Indicative fields: title, author(s), type, subject area, repository, language, publisher, publication date.
        
        \item \textit{Newspapers and Periodical/Reviews}: Allows documenting information about (old) newspapers and periodical/reviews. Indicative fields: title, type, subject area, author, language, editor, publisher, publication date.
        
        \item \textit{Oral History Sources}: Allows documenting information about oral history sources, like an oral testimony or interview. Indicative fields: title, subject area, description, language, interview date, interviewer, interviewee.
        
        \item \textit{Web Sources}: Allows documenting information about web sources providing historical information about one or more entities of interest. Indicative fields: URI, web page title, subject area, content language, text.

        \item \textit{Bibliography}: Allows documenting information about bibliographic references related to the project. Indicative fields: type, title, author(s), publisher, publication date/place, conference title, volume and issue number, language.

        \item \textit{Source Passages}: Allows documenting information about a specific source passage that provides important information for an entity of interest (e.g., an object transfer). Indicative fields: title, subject area, topic, origin (source or bibliography), source passage text, translation, commentary.
        
        \item \textit{Collection of Source Passages}: Allows grouping a set of source passages, e.g., based on an object, source, etc. Indicative fields: title, subject, short description, source passage(s).
        
        \item \textit{Researcher Comments}: Allows documenting information about research results, e.g., the findings of observing the inscriptions of an icon. Indicative fields: researcher, title, about (object, transfer, route, historical figure), description, date, based on (type of research), conclusion, property of analysis, outcome of analysis, method of analysis, date of analysis.

        \item \textit{Historical Figures}: Allows documenting information about historical persons, like a bishop, patriarch, etc. Indicative fields: name, role, service, birth place, ethnicity, life period, activity period, references.
        
        \item \textit{Collections}: Allows documenting information about collections of objects, e.g., museum collections. Indicative fields: code number, subject, originator of reference, description.
        
        \item \textit{Events}: Allows documenting information about historical events, such as a prince reception, an archbishop ordination, or the erection of a church. Indicative fields: name, time of event, location, description, references.
        
        \item \textit{Locations}: Allows documenting information about locations, such as cities, villages, monasteries, churches and museums. Indicative fields: name, location type, geopolitical hierarchy, coordinates.
        
        \item \textit{Persons}: Allows documenting information about persons (not historical), such as the researchers participating in the project, a photographer, etc. Indicative fields: name, name in native language, role, member of, description.
        
        \item \textit{Organisations}: Allows documenting information about organisations, such as museums, libraries, ephorates, etc. Indicative fields: name, type, pursuit (field), location, contact information, description.
        
        \item \textit{Digital Objects}: Allows documenting metadata information about the uploaded digital objects like photos. Indicative fields: title, type, short description, file, rights, creation date, creator.
\end{itemize}

\vspace{0.5mm} \noindent
{\bf Types of Documentation Fields.}
Each documentation field in \system\ has a particular type which specifies the type of value that it can receive. The supported types are the following:
\begin{itemize}
    \item \textit{Link to entity}. The user can select another entity that is documented in the system. The entity can belong to one or more (pre-defined) entity types. The fields \textit{Originator of Reference} (link to Organisation) and \textit{Current Location} (link to Location) of an object are examples of this field type.
    
    \item \textit{Link to vocabulary term}. The user can select a term from a static or dynamic vocabulary. A dynamic vocabulary allows users to directly create a new term (which is then added in the vocabulary), while a static vocabulary limits the options to a specific set of terms. An example of a static vocabulary is the \textit{Category} of an object and an example of a dynamic vocabulary is the \textit{Publisher Name} of a book. Both static and dynamic vocabularies can be managed through an administration page in \system. 
    
    \item \textit{Link to thesaurus term}. The user can select a term from a thesaurus of terms which is managed through the THEMAS thesaurus management system. THEMAS\footnote{\url{https://www.ics.forth.gr/isl/themas-thesaurus-management-system}} is an open source Web-based system for creating, managing and administering multi-faceted and multilingual thesauri according to the principles of ISO standards 25964-1 and 25964-2. THEMAS offers an API which allows its connection with \system.
    
    \item \textit{Unformatted free text}. The user can provide a piece of text that is usually small in size and that cannot be formatted. Examples of fields of such type are the fields \textit{Object Name} and \textit{Object Code} of the entity type Object.
    
    \item \textit{Formatted free text}. The user can provide a piece of text that is usually long in size and which can be formatted. The field \textit{Transfer Description} of an Object Transfer is an example of this type.
    
    \item \textit{Number}. The user can provide a numeric value, e.g., an integer number. The field \textit{Dimension Value} of an object measurement is an example of this type.
    
    \item \textit{Time expression}. The user can provide a date range in an accepted format relevant to the documentation of historical information, such as \textit{decade of 1970}, \textit{ca. 1920}, \textit{1st half 4th century}, \textit{1500 BCE}, \textit{3rd century - 5th century}).\footnote{The full list of the accepted time expressions is available at: \url{https://isl.ics.forth.gr/FeXML_ricontrans/HelpPage_en.html}} 
    Restricting the accepted value types of a date range is important for enabling comparisons and effective data exploration. An example of a documentation field of this type is the field \textit{Creation/Production Date} of an object.
    
    \item \textit{Location coordinates}. The user can select a point or polygon on a map and the field will be automatically filled with the corresponding coordinates. 
    
    \item \textit{Location ID}. The system offers the capability to query external geolocation services, in particular Getty Thesaurus of Geographic Names (TGN)\footnote{\url{http://www.getty.edu/research/tools/vocabularies/tgn/}} or Geonames\footnote{\url{https://www.geonames.org/}}, and get the unique ID and the coordinates of a location.
    
    \item \textit{Digital file(s)}. The user can upload one or more digital files of a given file type, e.g., image or document. 

\end{itemize}

A field can also be defined as \sq{multiple}, which means that the user can create multiple instances of it. In case a multiple field is not a leaf, the whole structure (having the field as root) is duplicated. An example of such a multiple field is the \textit{Dimension} field of the object schema (shown in Fig.~\ref{fig:schemaExample}), allowing to add multiple dimensions for a measurement, such as height and width. 

The total number of documentation fields in all entity types that link to other entities is currently 158, showcasing the high connectivity of the documented entities.
The number of distinct vocabularies is currently 113, while the number of documentation fields in all entity types that link to vocabulary terms is 244, with objects having the highest number of such fields (123, in total).
Also, there are two fields that link to a thesaurus in THEMAS: the field \textit{object kind} (of objects and object transfers) and the field \textit{topic} (of objects and source passages).

\subsection{Data Transformation (for Knowledge Graph Production)}
\label{subsec:transformation}

For transforming the documented data to a rich semantic network of interconnected data (an RDF knowledge graph), we make use of the X3ML framework and the X3ML mapping definition language \cite{marketakis2017x3ml}, a declarative, XML-based language that supports the cognitive process of schema mapping definition. X3ML separates schema mappings from the generation of proper resource identifiers (URIs), so it distinguishes between activities carried out by the domain experts and data engineers, who know the data, from activities carried out by the IT experts who implement data transformation. 

Given our target domain ontology (CIDOC-CRM), we need to create one mapping file for each schema of \system, i.e., for each entity type. 
In general, the definition of the mappings from the source schemas to the target ontology is a time-consuming process that can require many revisions as long as the data engineer better understands the data or changes are made to the schemas of the entity types. This process is supported by 3M Editor \cite{marketakis2017x3ml}, an X3ML mapping management system suitable for creating and handling the mapping files. It offers a user interface and a variety of actions that help experts manage their schema mappings collaboratively. 
For technical details about the data transformation process and the tools X3ML and 3M Editor, the reader can refer to \cite{marketakis2017x3ml}. 

The hierarchies of terms created in THEMAS are represented in RDF using SKOS (Simple Knowledge Organization System\footnote{\url{https://www.w3.org/TR/skos-reference/}}), while the vocabularies maintained in \system\ are represented using the class \textit{E55~Type} of CIDOC-CRM. 

The transformation of the data to a CIDOC-CRM compliant semantic network increases their value and their long term validity, facilitates integration with other CIDOC-CRM compliant datasets, and enables their advanced querying, analysis and exploration (more about the latter in Sect.~\ref{sec:lessonsLearnedAndFuture}).

\section{User Interface and Usage Statistics}
\label{sec:UI}
We present the web interface of \system\ (Sect.~\ref{subsec:userInterface}), describe how the documentation process is performed (Sect.~\ref{subsec:documentation}), and provide usage statistics (Sect.~\ref{subsec:stats}). 

\subsection{The \system\ Web Interface}
\label{subsec:userInterface}

\begin{figure}
    \vspace{-3mm}
    \centering
    \includegraphics[width=12.5cm]{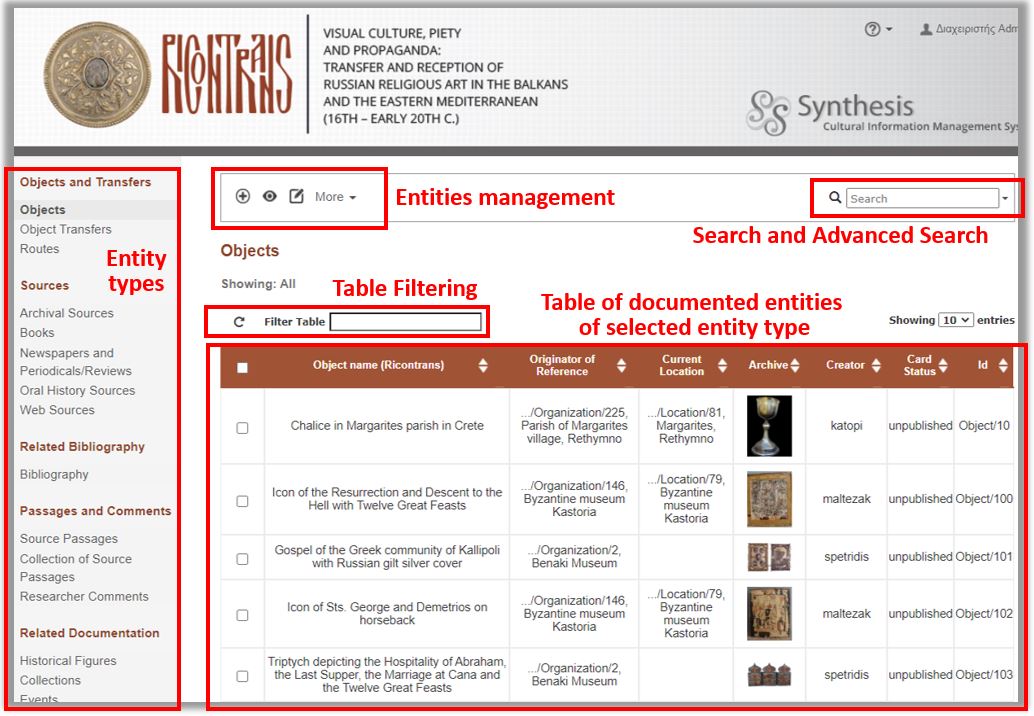} 
    \vspace{-6.5mm}
    \caption{The user interface of \system\, displaying the supported entity types and the table of documented entities of type \sq{Object}.}
    \label{fig:UI1}
    \vspace{-3.5mm}
\end{figure}

The interface of \system\ is Web-based and quite simple. After a successful user login, the homepage contains a left menu showing all the supported entity types, grouped in categories (Fig.~\ref{fig:UI1}). For each entity type, the user is shown a table with all entities that belong to the selected type and that are currently documented in \system, as shown in Fig.~\ref{fig:UI1} for the entity type \sq{Object}. 
For each entity, the table shows some basic information which is configurable for each different entity type. For example, for an entity of type Object the table shows its \textit{name}, its \textit{originator of reference} (the organisation responsible for the object), its \textit{current location}, an \textit{image}, the \textit{creator} of the documentation entity (i.e., the researcher who takes care of the object's documentation), the card \textit{status} (unpublished, pending, published; allows tracking the status of the entity's documentation card), and its \textit{ID} (automatically assigned by the system). 

The user can \textit{filter} the entities shown in the table by writing some text in an input field that exists above the table (cf.~Fig.~\ref{fig:UI1}). In this case, the table shows only those entities for which any of the characteristics shown in the table match the input text. 
Also, the system offers a \textit{search} functionality, which allows keyword-based searching within the entity's documentation fields, as well as an \textit{advanced search} functionality which allows searching based on values on specific fields, such as searching for object transfers having the value \sq{donation} as purpose of transfer and the date \sq{within 18th century} as the transfer date. Advanced search also provides the option to save a query in order to use it in the future (either from the same user or from other users).

From the web page of a particular entity type, the user has the following options: 
i)~create a new entity for documentation,
ii)~view the documentation card of one of the entities that appear in the table,
iii)~edit one of the entities (its documentation card), 
iv)~request for publishing one or more entities (which means that the documentation of these entities has been completed and no more editing is required),
v)~create a new version of an entity, 
vi)~view the existing versions of an entity, 
vii)~delete one or more entities,
viii)~create a copy of an entity (for documenting a similar entity),
ix)~give edit rights for one or more entities to another user account, 
x)~export the schema of the entity type,
xi)~export the data of one or more entities in XML or RDF format,
xii)~import an XML of a documented entity.

Regarding the export of data to RDF, the  exported data will be CIDOC-CRM compliant if there is an X3ML mapping file for the corresponding entity type. If there is no such mapping file, a naive (ontology-agnostic) schema is used for transforming the data to RDF.

\begin{figure}
    \vspace{-3mm}
    \centering
    \includegraphics[width=12.4cm]{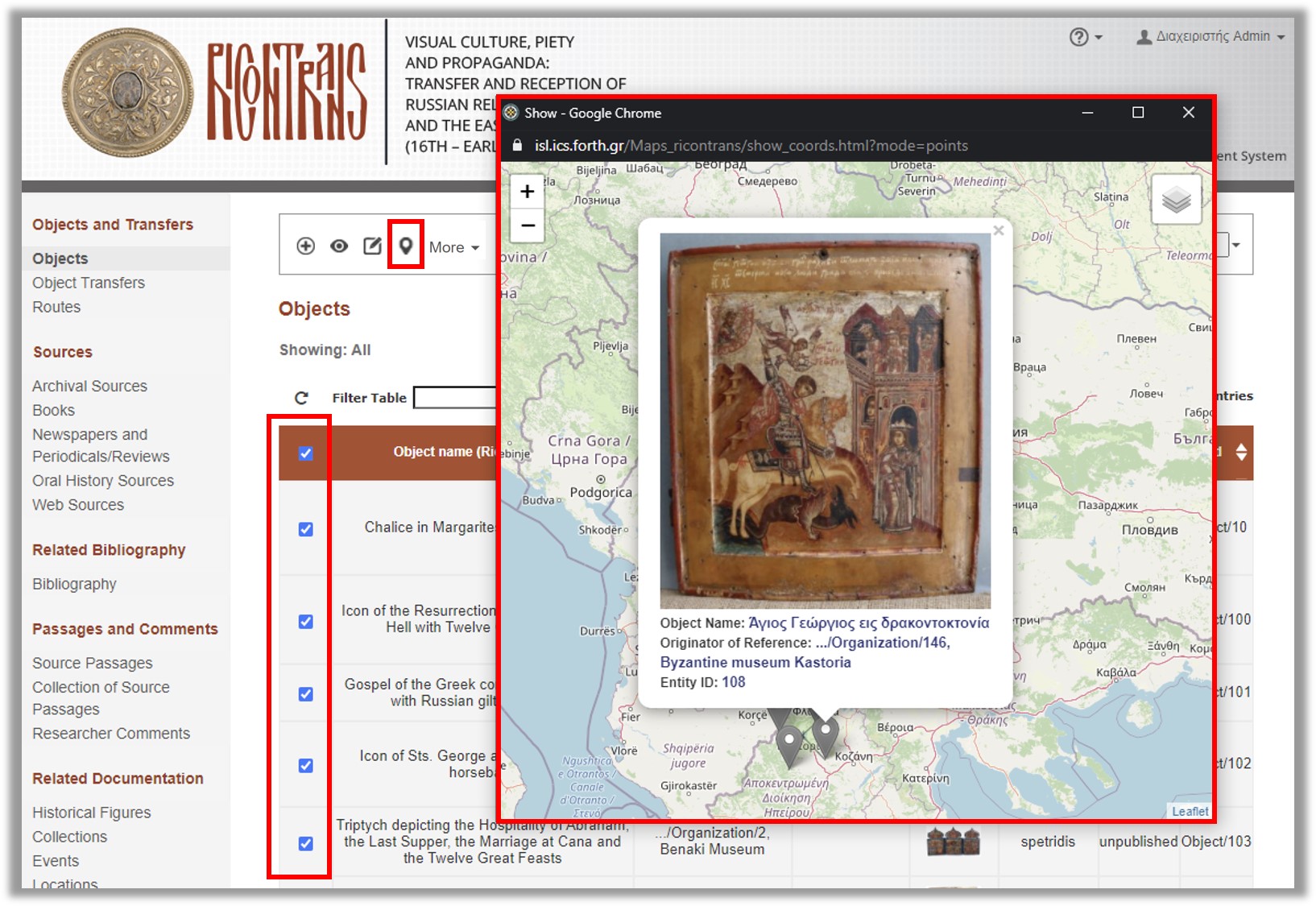} 
    \vspace{-6.5mm}
    \caption{Displaying a set of objects in a map.}
    \label{fig:UI_map}
    \vspace{-4mm}
\end{figure}

For certain entity types, the user can select one or more entities and display them on a map. In RICONTRANS, this option is currently available for the entity types \textit{Location}, \textit{Object} (showing the current location of the selected objects; as shown in Fig.~\ref{fig:UI_map}), \textit{Object Transfer} (showing lines connecting the starting and ending locations of the selected transfers), and \textit{Route} (showing sets of object transfers). The information to show for each point in the map is configurable per entity type.

\subsection{Entity documentation}
\label{subsec:documentation}

The documentation of entities is performed in a dedicated environment called FeXML, which communicates with \system\ and supports the creation and editing of XML documents. 
FeXML is activated when the user a)~creates a new entity for documentation, b)~selects to view one of the documented entities (its {\em documentation card}), or c)~selects to edit one of the documented entities (e.g., for continuing its documentation, correcting a field value, etc.). 

\begin{figure}
    \vspace{-4mm}
    \centering
    \includegraphics[width=12.7cm]{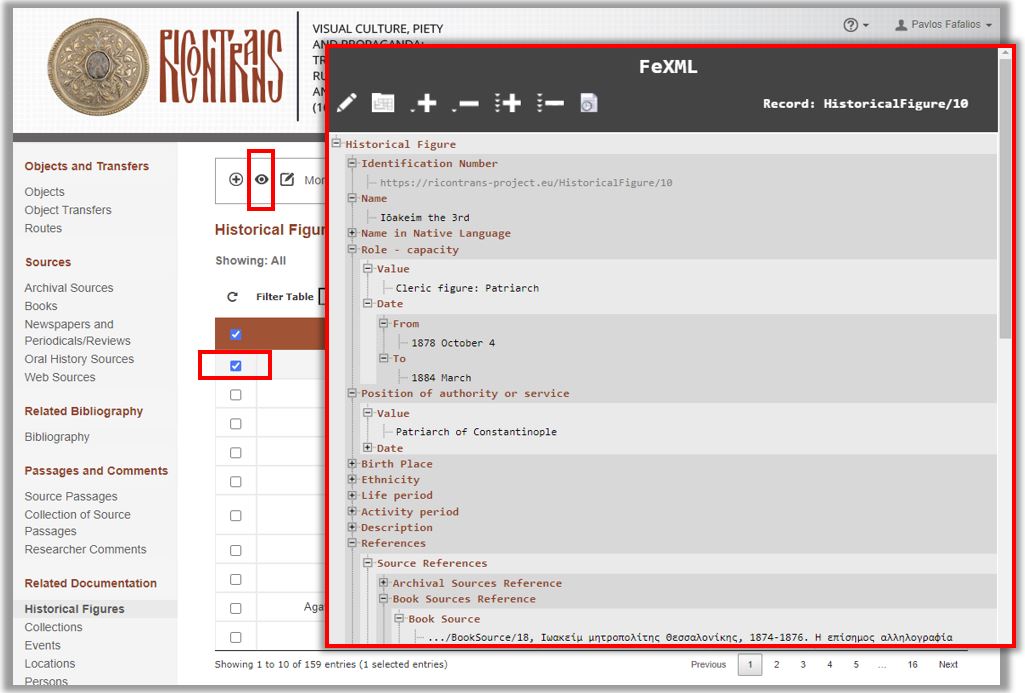} 
    \vspace{-6.5mm}
    \caption{Viewing the documentation card of an entity.}
    \label{fig:ui_fexml}
    \vspace{-3mm}
\end{figure}

Fig.~\ref{fig:ui_fexml} shows an example of the documentation card of an entity of type \sq{Historical Figure}, as shown in FeXML in view mode. 
The documented information is shown in a tree-like structure, where  the root of the tree is the name of the entity type and the leaves are the documentation fields. The user can expand or collapse fields on-demand, in order to facilitate its navigation to the documentation fields. By default, when the documentation card of an entity is viewed, FeXML shows as expanded only the filled fields. In the bottom of the documentation card, the user is also shown with all the entity's associations with other documented entities (those the entity references and those the entity is referenced by).

\begin{figure}
    \vspace{-3mm}
    \centering
    \includegraphics[width=12.7cm]{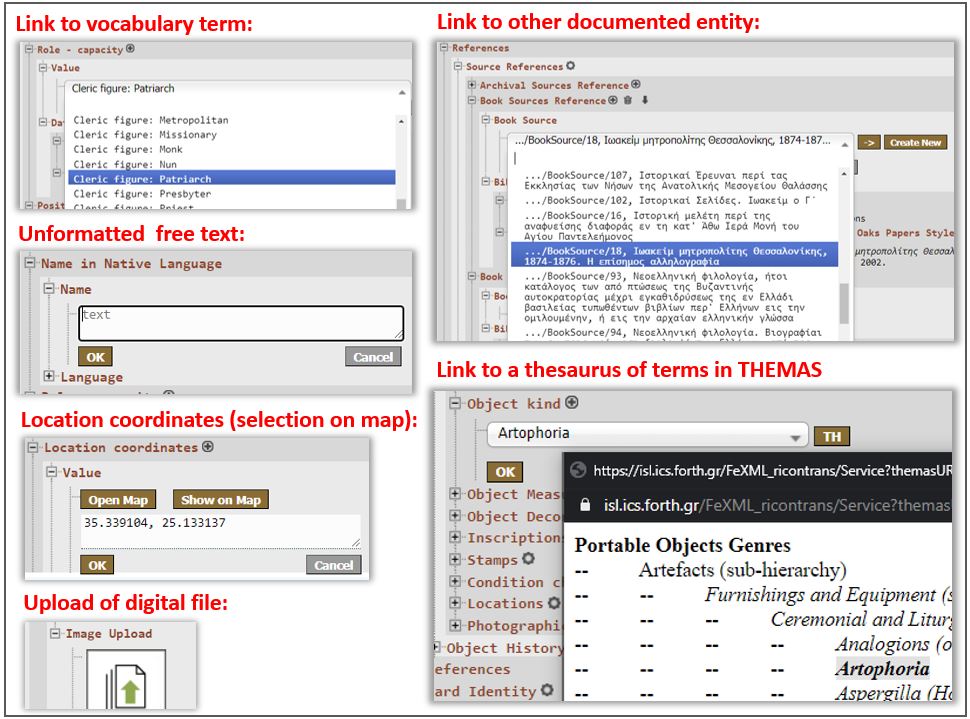} 
    \vspace{-6.5mm}
    \caption{Examples of documentation fields of different types.}
    \label{fig:ui_docFields}
    \vspace{-4mm}
\end{figure}

In edit mode, the user can start filling the available documentation fields. Fig.~\ref{fig:ui_docFields} shows examples on how the user can fill information for different types of fields. 
Also, there is a button on the top of the FeXML window which allows users to see all the accepted time expressions, as well as a button which opens an XML Map showing the full hierarchy of the fields (for facilitating navigation especially when the documentation card is very lengthy, like in the case of objects).

Finally, the management of the vocabularies is performed through a dedicated page accessible through the administration menu of \system. This menu option is only shown to user accounts that have the rights to edit the vocabularies. 
The user can select a vocabulary and \textit{add} new terms, \textit{edit} or \textit{delete} existing terms, as well as \textit{export} the whole vocabulary or \textit{import} a vocabulary from a text file. 

\subsection{Usage statistics}
\label{subsec:stats}
The system is currently being used by 40 users (historians, art historians, philologists, and students in these fields) belonging to 10 institutions in 5 countries. 
The current (as of July 2, 2021) number of documented entities per entity type is: 
1,089  objects,
368 object transfers,
93 routes,
150 archival sources,
45 books,
98 newspapers and periodicals/reviews,
3 oral history sources,
59 web sources,
309 bibliographic items,
533 source passages,
3 collections of source passages,
203 historical figures,
155 collections,
33 events,
491 locations,
98 persons,
394 organisations, and
1,138 digital objects.


\section{Lessons Learned and Future Data Exploitation}
\label{sec:lessonsLearnedAndFuture}

Collaborating with historians in the context of RICONTRANS gave us the unique chance to learn how domain experts in this area work, what their questions are, what data is important for them to document, and most importantly, what difficulties they face in terms of data management.  
Below we provide some lessons learned from this collaboration: 

\textit{Selection of entity types and design of schemas}. The decision on the entity types to support as well as on the documentation fields of each type requires extensive discussions between domain experts (historians) and data engineers. The challenge here is to find the best trade-off between documentation richness and usability, as well as convince domain experts that some additional entity types and documentation fields are required for long-term usefulness of the data and support of better data exploration services. For example, the inclusion of \sq{Source Passages} as a different entity type, and not as part of the sources, makes documentation a bit more complex but allows linking object transfers to specific source passages and thus supports answering queries such as: \textit{\q{Give me source passages that talk about  transfers of icons from Russia to monasteries in Mount Athos.}} Moreover, it makes the sources independent of the source passages that are of interest in RICONTRANS (thus, in future one may link the same source to other source passages).
Similarly, it is much simpler to record the dimensions of an object in a single text field (e.g., \q{15cm x 20cm}) than breaking it to 3 fields (\textit{property}, \textit{value}, \textit{unit}). However, the former makes very difficult, if not impossible, to make comparisons between the size of objects. 

\textit{Controlling the dynamic vocabularies}. The documentation fields of type \sq{dynamic vocabulary} allow users to create a new term which is then added in the vocabulary and is available for selection by other users. The problem here is that some users do not carefully check the list to see if the desired term already appears in the vocabulary and create a new term. If the new term already exists, the user is informed and then can select it from the list. However, if the user gives a different name for a term which already exists in the vocabulary, then the new term is included in the vocabulary. This results in vocabularies containing multiple terms that refer to the same concept, making their future exploitation difficult. For instance, in a search system one might search for all events of a particular category (type) but do not get back all relevant events because the same category has been included in the vocabulary multiple times with different names/labels. Thereby, there is a need for curating the dynamic vocabularies frequently, which is time consuming and may require the curator to contact other users for understanding the meaning of some specific terms.

\textit{Understanding the tree-like structure of the documentation fields}. Many users seem to get confused with the tree-like structure of the documentation fields, which sometimes results in data entry errors. For example, the same field \sq{Name} exists in multiple positions in the hierarchy of the fields, e.g., \q{Object/.../Object History/Event/Name}, \q{Object/.../Object History/Use/Name}, \q{Object/.../ Object History/Acquisition/Name}. To solve this problem we renamed many fields so that their meaning is clear (e.g., from \q{Name} to \q{Event Name}).

\vspace{2mm}
\noindent 
{\bf Data dissemination and exploitation.} 
The project is currently in the data entry phase where users gather and document information about entities of interest.
Apart from the study of the documented data for historical research, part of the data will be made publicly available.
The current plan for dissemination and exploitation comprises three main services: 
1)~\textit{map visualization}, 
2)~\textit{data publication}, and 
3)~\textit{semantic network exploration}.

{\em Map visualisation}. Part of the data, such as objects and object transfers, will be visualised on a Web-accessible map application. This will allow historians and other interested parties to explore the data in an interactive environment and learn about the historical routes of several religious artefacts brought to the Balkans the period 16th to 20th centuries.

{\em Data publication}. Part of the data, such an object of high interest and its transfers, will be presented in web pages. The information shown in the web pages will be directly linked to the data in \system, which means that updates in \system\ will be directly reflected in the web pages.

{\em Semantic network exploration}. The CIDOC-CRM compliant semantic network (RDF graph), derived from the data transformation process (cf.~Sect.~\ref{subsec:transformation}), can support the advanced exploration of the documented data and the answer of complex information needs. 
The user-friendly exploration of such a network can be performed through two main general access methods: 
i) \textit{keyword search}, where the user submits a free text query and gets back a ranked list of results that are relevant to the query terms (e.g., \cite{kadilierakis2020keyword,nikas2020keyword}),
and 
ii) \textit{interactive access}, where the user explores the data through intuitive interactions with a data access system, e.g., using a \textit{faceted search} interface  \cite{tzitzikas2017faceted} or \textit{assistive query building} (like in \cite{oldman2018reshaping} and \cite{kritsotakis2018assistive}).  
Our plan is to make use of an assistive query building interface which will support users in finding answers to information needs that require exploiting the rich associations among the entities and their characteristics, such as \textit{\q{find me objects of type \sq{icon} transferred from Russia to monasteries in Greece as a donation}}, or \textit{\q{find me sources passages that talk about donations of icons transferred to Greece before the 18th century.}}

\section{Conclusion}
\label{sec:conclusion}

We have presented the use of the \system\ system for data documentation and management in the context of a large-scale research project in the field of History of Art, called RICONTRANS. 
The system is Web-based, collaborative and makes use of established standards (CIDOC-CRM, RDF) for information documentation and publication, facilitating data integration and reuse, and focusing on the production of data with high value, long term validity and longevity.

\system\ provides full-fledged support for the complete knowledge production life-cycle in historical research. It is currently used by a large number of historians for the documentation of data about religious artefacts, their transfers, sources of information like archival and oral history sources, and other involved entities, such as  historical figures and locations. 

An interesting direction for future work is the application of information extraction techniques that can facilitate or accelerate data entry, such as the use of named entity extraction \cite{mountantonakis2020lodsyndesisie} for semi-automatically filling the \sq{referenced information} fields of source passages (referenced persons, locations, dates, etc.).

\subsection*{Acknowledgements}
This work has received funding from the European Union's Horizon 2020 research and innovation programme under 
i) the Marie Sklodowska-Curie grant agreement No 890861 (Project \q{ReKnow}), 
and ii) the European Research Council (ERC) grant agreement No 818791 (Project \q{RICONTRANS}).

\bibliographystyle{splncs04}
\bibliography{SynthesisRicon__MAIN}

\end{document}